\documentclass[amsmath,comment,amssymb,twocolumn,showpacs,pre,aps]{revtex4}

\usepackage{graphicx}

\usepackage{amssymb}

\usepackage{epsfig}

\usepackage{dcolumn}

\usepackage{bm}

\usepackage{color}

\newcommand{\be}{\begin{equation}}

\newcommand{\ee}{\end{equation}}

\newcommand{\bea}{\begin{eqnarray}}

\newcommand{\eea}{\end{eqnarray}}

\newcommand{\nn}{\nonumber }

\begin{document}

\title{Enhancement of chaotic subdiffusion in disordered ladders with synthetic gauge fields}

\author{Xiaoquan Yu and Sergej Flach}

\affiliation{New Zealand Institute for Advanced Study, Centre for Theoretical Chemistry and Physics,Massey University, Auckland 0745, New Zealand }

\date{\today}

\begin{abstract}
We study spreading wave packets in a disordered
nonlinear ladder with broken time-reversal symmetry induced by synthetic gauge fields.
The model describes the dynamics of interacting bosons in a disordered and driven optical ladder within a mean-field approximation.
The second moment of the wave packet
$m_{2} = g  t^{\alpha}$ grows subdiffusively with the universal exponent $\alpha \simeq 1/3$ similar to the 
time-reversal case. However, the prefactor $g$ is strongly modified by the field strength and shows a non-monotonic dependence. For a weak field, the prefactor increases 
since time-reversal enhanced backscattering is suppressed.
For strong fields the spectrum of the linear wave equation reduces the localization length through the formation of gaps and narrow bands. Consequently the prefactor for the subdiffusive spreading law is suppressed.
\end{abstract}

\maketitle

\section{Introduction}
The spreading of wave packets in disordered nonlinear lattices has regained a lot of 
interest recently. Many experiments were performed with ultracold atomic condensates in random optical potentials~\cite{BEC}.
Due to disorder, linear waves which correspond to noninteracting condensates will eventually stop spreading 
and exponentially localize in low dimensions~\cite{Anderson}. 
Mean-field treated two-body interactions lead to cubic nonlinear terms in the wave equations. They induce overlaps between the normal modes of the linear wave equation,
and ultimately lead to chaotic dynamics. The subsequent decoherence of phases of the normal modes breaks localization through incoherent spreading
\cite{brokenlocalization,Sergej}.
Numerical studies on wave packet spreading in several one-dimensional nonlinear 
disordered lattice models show that the wave packet exhibits a subdiffusive behavior for weak nonlinearity; 
namely, the second moment $m_2$ of a wave packet grows as $m_2 =  g t^{\alpha}$ with the universal exponent $\alpha \simeq 1/3$~\cite{Sergej}.
Several issues remain under debate, especially 
the asymptotic behavior of the second moment in the long time limit~\cite{Benettin,Wang,Fishman,Basko,Mulansky}.  

Previous studies focused only on time-reversal symmetry (TRS) cases. 
Here we consider the situation of broken TRS.
For charged particles, TRS can be broken by turning on magnetic fields. For neutral particles, like bosons, 
we can introduce synthetic gauge fields~\cite{sytheticgauge}.
The advantage of synthetic gauge fields on optical lattices is that strong fields, where the flux per lattice
cell can take any value between zero and 2$\pi$, can be realized with lasers~\cite{sytheticgauge}. 
This has to be contrasted to the case of electrons in metals~\cite{footnoteelectron}. 
Therefore synthetic gauge fields offer an opportunity for experimental studies on many lattice models under strong magnetic fields~\cite{Bloch}.

In the presence of magnetic fields, one may expect that wave spreading slows down due to 
the Lorentz force, which tends to localize an excitation. On the other hand, the magnetic field breaks TRS and therefore enhances the 
localization length in a disordered system. Moreover the magnetic field can also change the band structure of the corresponding linear and ordered system.

In this work, we study the influence of synthetic gauge fields on the 
spreading of nonlinear waves. 
It turns out that each of the effects mentioned above exhibits its dominant role in different parameter and time regimes.  

\section{Model}
\begin{figure}[h]
\includegraphics[width=2.6in]{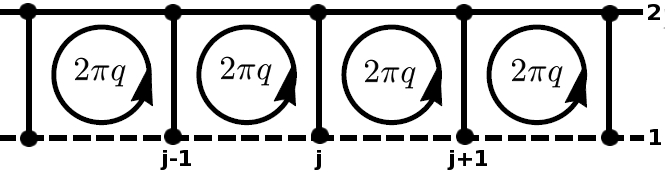}
\vspace{.1cm}
\caption{Two-leg ladder lattice exposed to synthetic gauge fields. A uniform synthetic magnetic flux per plaquette is $2\pi q$. 
The gauge is chosen to reside only on chain 1. Solid lines correspond to hopping $t=1$ and dashed lines correspond to complex hopping. 
}
\label{f:twolegladder}
\end{figure}

We consider the discrete nonlinear Schr\"{o}dinger model on a two-leg ladder 
lattice (Fig.~\ref{f:twolegladder}) with complex hopping terms:
\begin{align}
\label{NonlinearHamiltonian}
H=&-t\sum_{l}\left(e^{-i2\pi q}\psi^{\ast}_{1,l}\psi_{1,l+1}+\psi^{\ast}_{2,l}\psi_{2,l+1}+{\rm h.c.}\right) \nn\\
&-t\sum_{l}\left(\psi^{\ast}_{1,l}\psi_{2,l}+{\rm h.c.}\right)\nn\\
&+\sum_{\nu=1,2}\sum_{l} |\psi_{\nu,l}|^{2} \left(\epsilon_{\nu,l} + \beta |\psi_{\nu,l}|^{2}\right).
\end{align}
Here $\nu\in{1,2}$ is the index labeling the two legs. $\psi_{\nu,l}$ is a complex field which quantifies  the order parameter of the atomic condensation. 
$\epsilon_{\nu,l}$ is a quenched, random uncorrelated on site potential taken to be uniformly distributed, $\epsilon_{\nu,l} \in[-\frac{W}{2},\frac{W}{2}]$. 
$t$ is the hopping strength and $\beta \geq 0$ is the nonlinearity parameter which is derived from the two-body interaction between atoms and is proportional to
the scattering length. Note that
the wave function $\psi_{\nu,l}$ is dimensionless. We measure energy and time in units of $t$ and $1/t$, respectively, and set $t=1$ without loss of generality. 
The uniform synthetic magnetic flux per plaquette is $2\pi q$. 
Here we choose the gauge such that the wave function only gains a phase $2\pi q$ 
when moving one lattice spacing along the chain leg $\nu=1$. 

It is convenient to introduce the two component wave function
\begin{eqnarray}
\begin{array}{cc}
                      \Psi_{l} \\
                      
                     \end{array} \equiv
 \left(\begin{array}{cc}
                       \psi_{1,l} \\
                       \psi_{2,l}
                     \end{array}\right),
\end{eqnarray} 
such that the Schr\"{o}dinger equation associated with the Hamiltonian~(\ref{NonlinearHamiltonian}) can be written as
\begin{eqnarray}
\label{EOM}
i\dot{\Psi}_{l}(\tau)={\rm D}_{l}\Psi_{l}-{\rm J}\Psi_{l+1}-{\rm J}^{-1}\Psi_{l-1}\nn\\
+\beta\sum_{\nu=1,2}({\rm I}_{\nu}\Psi_{l})({\rm I}_{\nu}\Psi_{l})^{\dagger}\Psi_{l},
\end{eqnarray}
where $({\rm I}_{\nu=1,2})_{ij}=\delta_{i,\nu}\delta_{j,\nu}$, and the $2\times2$ matrices $\rm J$ and ${\rm D}_{l}$ take the form 
\begin{eqnarray}
   {\rm J}=\left(\begin{array}{cc}
    e^{-i2 \pi q} & 0 \\
    0 & 1
  \end{array}\right), \quad
   {\rm D}_{l}=\left(\begin{array}{cc}
    \epsilon_{1l} & -1 \\
    -1 & \epsilon_{2l}
  \end{array}\right).
\end{eqnarray}
The eigenvalue problem of the linear part of Hamiltonian~(\ref{NonlinearHamiltonian}) is 
\begin{eqnarray}
\label{eigenfunction}
E_{\mu} \Phi_{\mu,l}=-{\rm J}\Phi_{\mu,l+1}-{\rm J}^{-1}\Phi_{\mu,l-1}+{\rm D}_{l} \Phi_{\mu,l},
\end{eqnarray}
where 
\begin{eqnarray}
\begin{array}{cc}
                      \Phi_{\mu,l} \\
                      
                     \end{array} \equiv
 \left(\begin{array}{cc}
                       \phi^{1}_{\mu,l} \\
                       \phi^{2}_{\mu,l}
                     \end{array}\right)
\end{eqnarray} 
is the $\mu$th eigenmode and $E_{\mu}$ is the corresponding eigenvalue.
Using the expansion $\Psi_{l}(t)=\sum_{\tilde{\mu}}c_{\tilde{\mu}}(t)\Phi_{\tilde{\mu},l}$ in Eq.~(\ref{EOM}), we obtain 
the equations of motion for the normal mode amplitudes
\begin{eqnarray}
\label{NLEOM}
i\dot{c}_{\mu}=E_{\mu}c_{\mu}+\beta\sum_{\mu_{1},\mu_{2},\mu_{3}} I_{\mu,\mu_{1},\mu_{2},\mu_{3}} c_{\mu_{1}}c^{\ast}_{\mu_{2}}c_{\mu_{3}}  
\end{eqnarray}
with 
\begin{eqnarray}
\label{OI}
I_{\mu,\mu_{1},\mu_{2},\mu_{3}}\equiv\sum_{l}\sum_{\nu=1,2}\Phi^{\dagger}_{\mu,l}{\rm I}_{\nu}\Phi_{\mu_{1},l}\Phi^{\dagger}_{\mu_{2},l}{\rm I}_{\nu}\Phi_{\mu_{3},l}.
\end{eqnarray}
The overlap integrals $I_{\mu,\mu_{1},\mu_{2},\mu_{3}}$ are random variables and their distribution plays a crucial role in the nonlinear wave packet spreading. 

\section{Linear equation properties}

\subsection{Clean system}
\begin{figure}[h]
\includegraphics[width=1.5in]{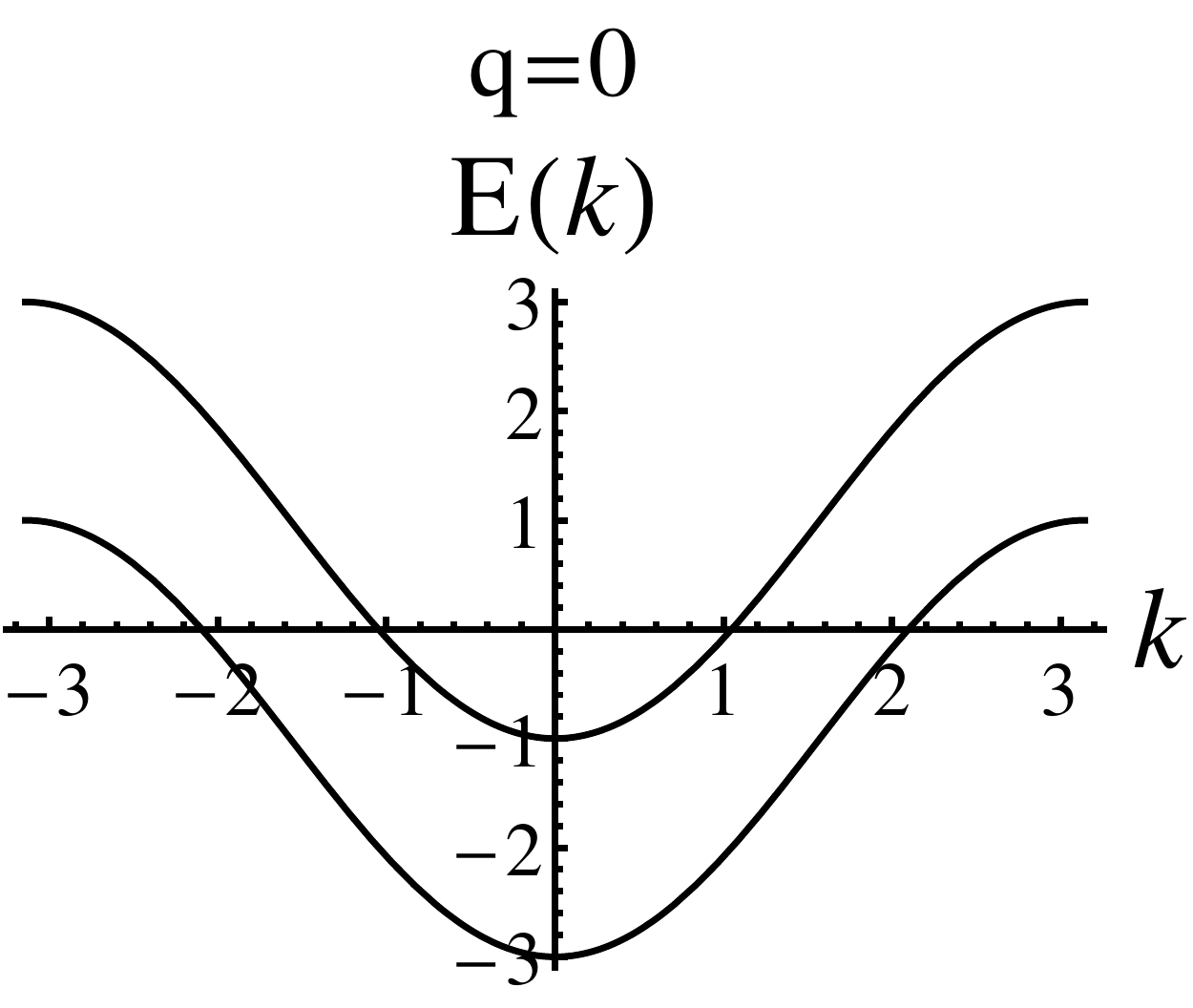}
\qquad
\includegraphics[width=1.5in]{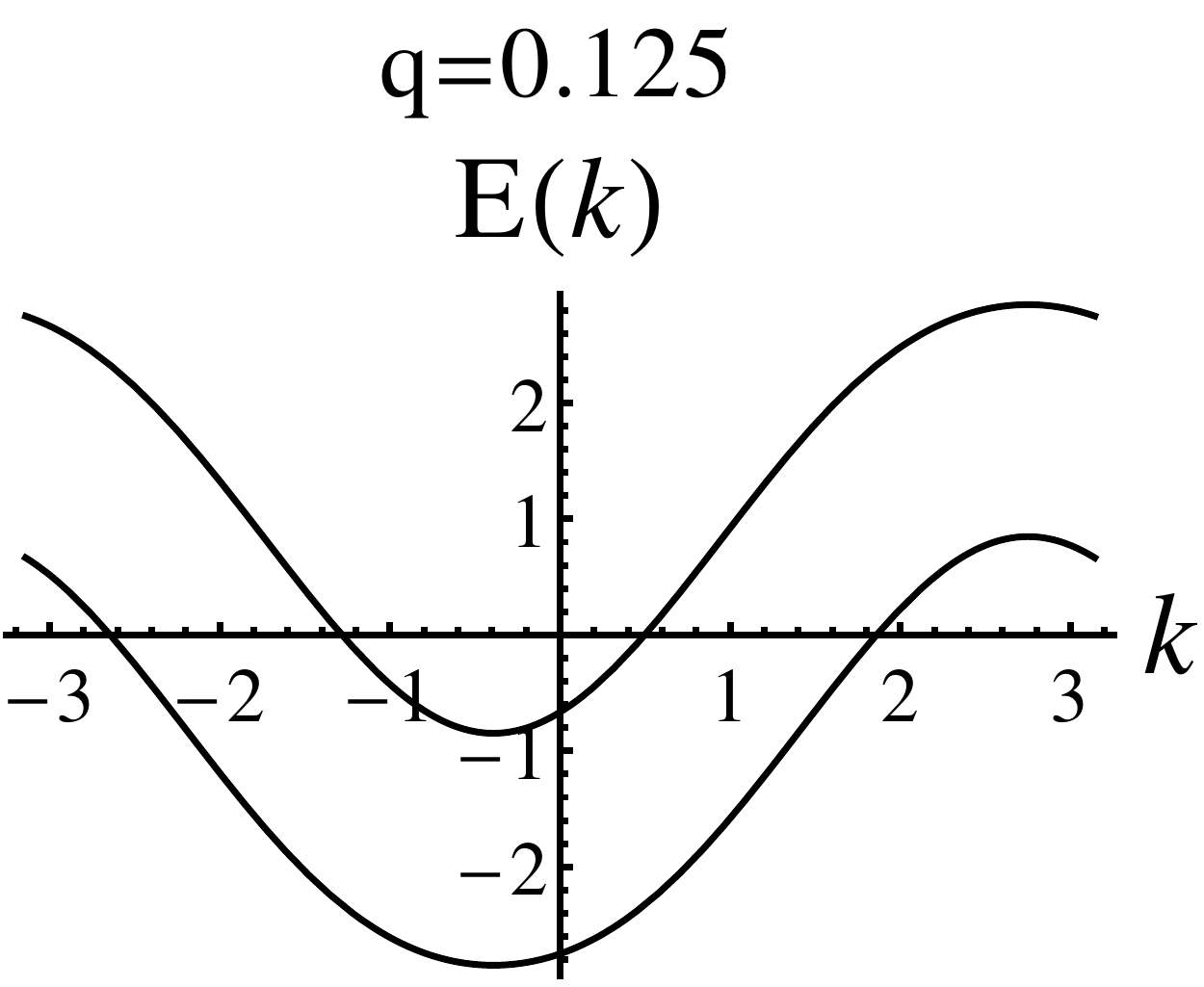}
\qquad
\includegraphics[width=1.5in]{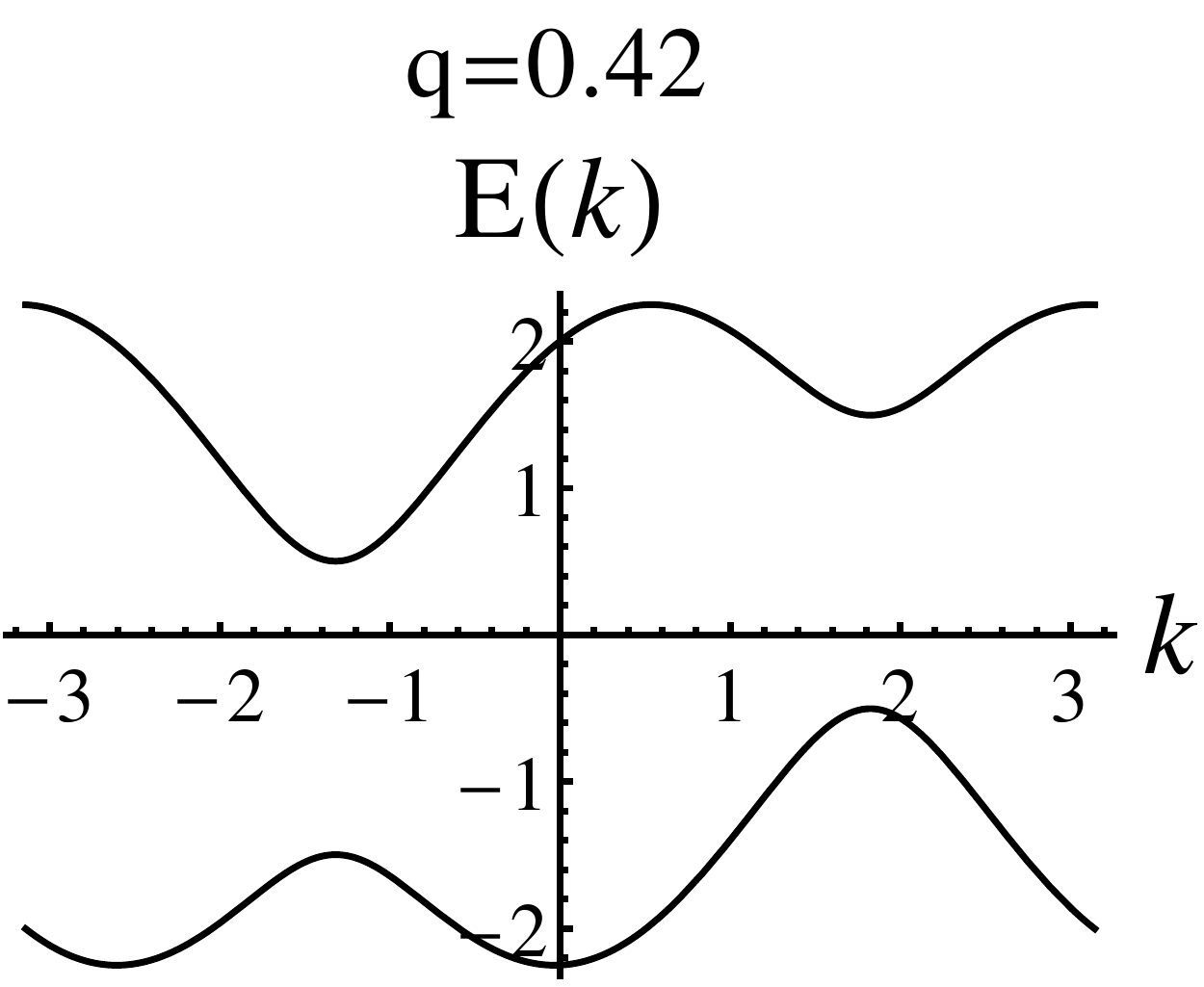}
\qquad
\includegraphics[width=1.5in]{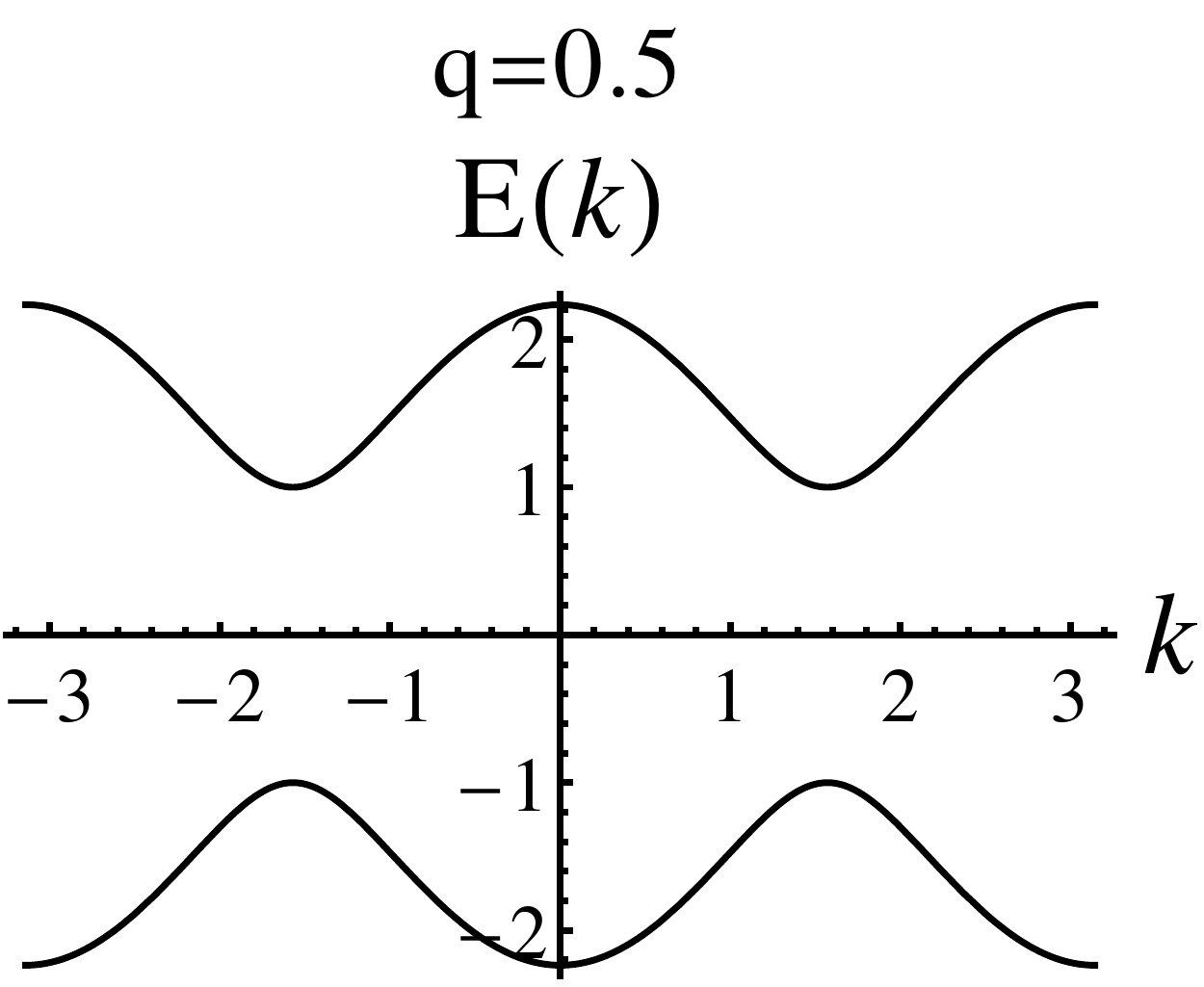}
\vspace{.1cm}
\caption{The band structure of the corresponding linear and clean model of Eq.~(\ref{NonlinearHamiltonian}) for various magnetic fluxes: 
$q=0,\ 0.125,\ 0.42,$ and $0.5$. The two bands open a gap when $q>q_c\simeq 0.34$. The gap is largest for $q=1/2$. At the same time, the bands become narrower.
}
\label{f:bandstructure}
\end{figure}

We first study the linear ($\beta=0$) and clean ($\epsilon_{\nu,j}=0$) case of Eq.~(\ref{NonlinearHamiltonian}).  
Let the length of the ladder be $L$. Under the gauge chosen and imposing periodic boundary conditions $\psi_{\nu,1}=\psi_{\nu,L}$,
the Hamiltonian~(\ref{NonlinearHamiltonian})
has lattice translation invariance along the horizontal direction and therefore can be easily diagonalized in the momentum space~\cite{footnotegaugefixing,Zak}. The eigenvalue spectrum reads
\begin{eqnarray}
\label{spectrum}
E_{\pm}(k)&=&-\cos(k)-\cos(2\pi q + k)\nn\\
&&\pm \sqrt{\left[\cos(k)-\cos(2\pi q +k)\right]^2+1},
\end{eqnarray}
where the momentum $k$ is varying in the first Brillouin zone $k \in [-\pi,\pi]$. 
Fig.~\ref{f:bandstructure} shows the band structure for various magnetic fluxes. Interestingly the two bands  
open a gap when the flux exceeds a certain critical value $q_{c}\simeq 0.34$, which is determined by the    
conditions $E_{\pm}(k)=0$ and $\frac{\partial E_{\pm}(k)}{\partial k}=0$. 
This behavior can be expected in a multibands system, since a magnetic field tends to reduce the kinetic energy of transnational motion and therefore
the band width of a single band. A magnetic field in a system with few bands will therefore flatten each band and eventually lead to the appearance of new gaps.

\subsection{Disorder}
For $\beta=0$, in the presence of disorder, all the normal modes are localized. 
The localization properties have been studied in details in the weak disorder limit ($W\ll 1$) in the absence of magnetic fields~\cite{Hongyi}.
Here we are interested in the effect of magnetic fields on the localization length in the regime of moderate disorder ($W=4$). 
We evaluate the localization length as a function of energy via the transfer matrix method for various magnetic fluxes.

Eq.~(\ref{eigenfunction}) can be rewritten in a recursive relation form (we ignore the subscript $\mu$ here)
\begin{eqnarray}
\left(\begin{array}{c}
  \Phi_{l+1} \\
  \Phi_{l}
\end{array}\right)={\rm T}_{l}(E)\left(\begin{array}{c}
                           \Phi_{l} \\
                           \Phi_{l-1}
                         \end{array}\right),
\end{eqnarray}
where the transfer matrix ${\rm T}_{l}(E)$ is given by
\begin{eqnarray}
\label{transfermatrix}
{\rm T}_{l}(E)&=&
\left(\begin{array}{cc}
 {\rm J}^{-1} \left({\rm D}_{l}-E\right)  & -({\rm J}^{-1})^{2} \\
  {\rm I} &  0
\end{array}\right)
\end{eqnarray}
with ${\rm I}_{i,j}=\delta_{i,j}$.

\begin{figure}[h]
\includegraphics[width=2.8in]{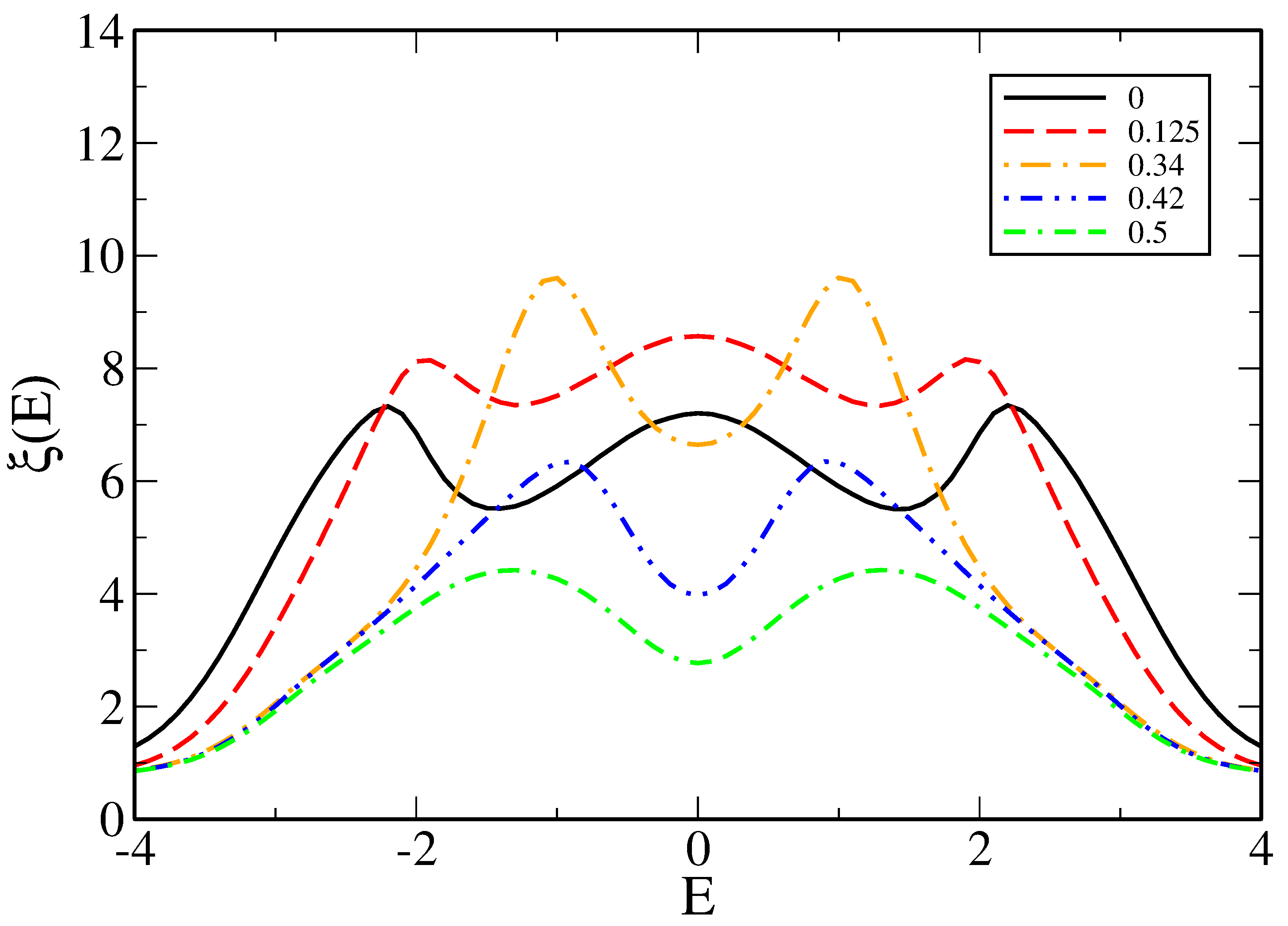}
\vspace{.1cm}
\caption{(Color online) Localization length as a function of energy with disorder strength $W=4$ for various magnetic fluxes: $q=0,\ 0.125,\ 0.34,\ 0.42,$ and $0.5$.
The system size $L=10^7$. The localization length is enhanced almost in the whole energy band when $0<q<q_c$, while it is reduced for 
$q_c < q \leq 1/2$ with $q_c\simeq 0.34$.}
\label{f:localizationlength}
\end{figure}

The localization length can be extracted from the matrix product 
${\rm M}_L={\rm T}_{L}(E)...{\rm T}_{2}(E){\rm T}_{1}(E)$.
The Lyapunov exponents associated with ${\rm M}_{L}$ are given by
$\gamma_i=\lim_{L\rightarrow \infty} \frac{1}{2L}\log \lambda_{i}$,
where $\{ \lambda_{i} \}$ are the four eigenvalues of ${\rm M}^{T}_{L}{\rm M}_{L}$. 
The localization length $\xi(E)$ is obtained by the inverse of the smallest positive Lyapunov exponent.
Note that in the presence of magnetic fields, the Lyapunov exponents do not appear in pairs with 
opposite sign,  as opposed to the TRS case.
We evaluate the Lyapunov exponents with an efficient numerical reorthogonalization method~\cite{reorthogonalization}.
Fig.~(\ref{f:localizationlength}) shows the localization length as a function of energy for various magnetic fluxes.

\begin{figure}[h]
\includegraphics[width=2.8in]{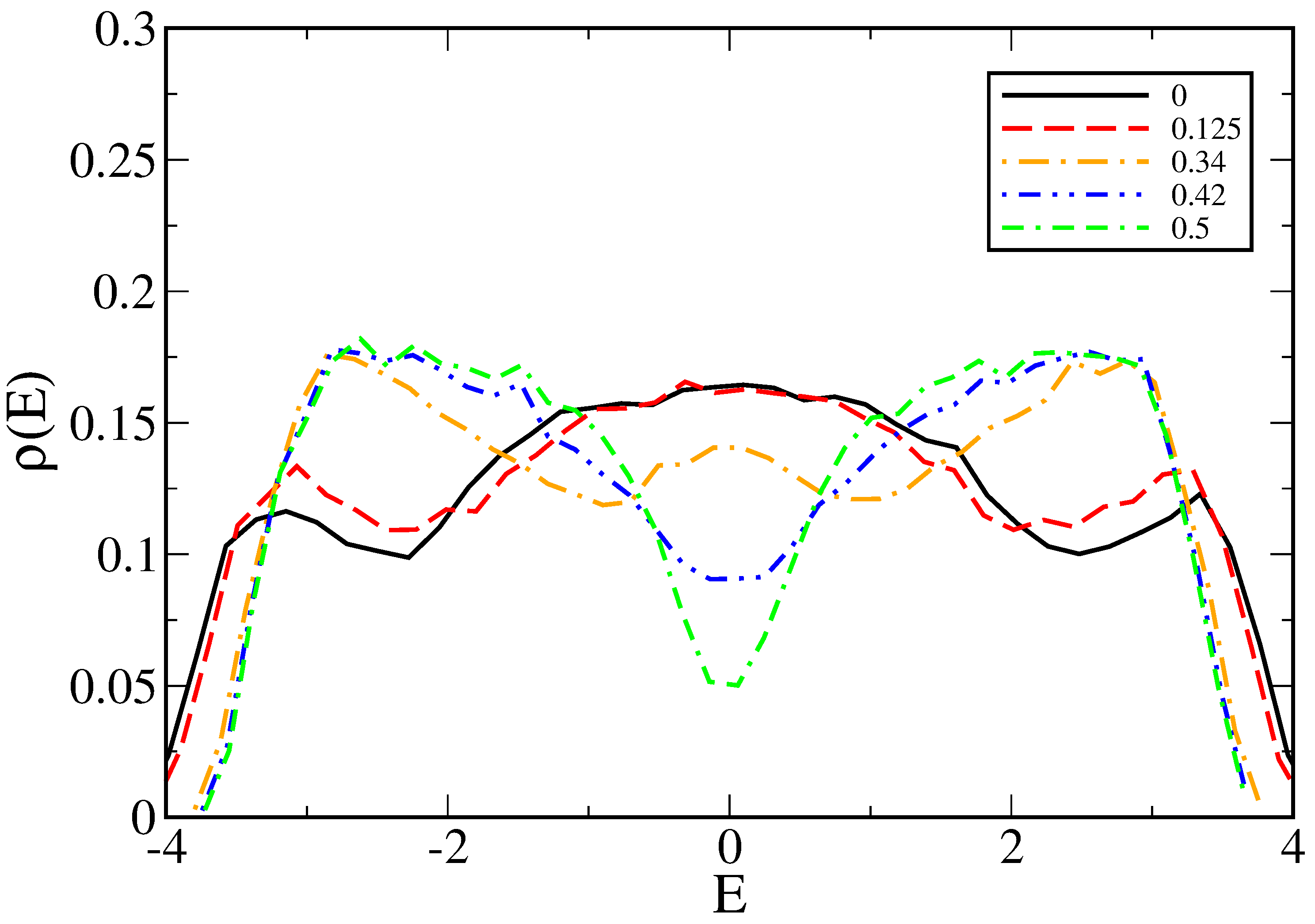}
\vspace{.1cm}
\caption{(Color online) Density of states averaged over 20 disorder realizations with disorder strength $W=4$ for various magnetic fluxes: $q=0,\ 0.125,\ 0.34,\ 0.42,$ and $0.5$.
The density of states opens a pseudogap around the center of the band when $q\sim q_c$ with $q_c\simeq 0.34$.}
\label{f:densityofstates}
\end{figure}

The magnetic field has two effects here.
First it breaks TRS, and second it modifies the band structure of the clean system and therefore the density of states. 
The density of states is defined as $\rho(E)=\frac{1}{N}\sum_{\mu}\langle\delta(E-E_{\mu})\rangle$,
where $N$ is the total number of states and the angular brackets
denote the averaging over the on site random potentials $\{\epsilon_{\nu,l}\}$.
  
For small magnetic fluxes ($0<q < q_c$), the localization length is enhanced in almost the whole energy band.
This is a well known phenomenon of the two-dimensional Anderson model in the presence of magnetic fields~\cite{weaklocalization}. 
It is due to the fact that magnetic fields, which destroy constructive interference by breaking TRS,
reduce the return probability and therefore enhance the localization length. 
  
Large magnetic fluxes ($q > q_c$) open an energy gap around $E=0$ in the corresponding clean system (see Fig.~\ref{f:bandstructure}). 
Therefore the density of states opens a pseudogap around $E=0$ (see Fig.\ref{f:densityofstates}).  
States in that pseudogap are similar to states in the Lifshitz tails in the band edges and 
the localization length of these states is shortened.     
At the same time the two bands of the clean system have a smaller width. This reduction of kinetic energy 
leads to a reduction of the localization length at energies away from the pseudogap region, where the density of states is also enhanced. 
Indeed, in one-dimensional
systems the localization length is proportional to the mean free path, and the enhanced density of states 
reduces the mean free path and therefore the localization length. 
At the largest value of the magnetic flux $q=1/2$, TRS is restored. The restoring of this symmetry
is another factor which leads to a suppression of the localization length in the whole energy range.

\section{Nonlinear wave packet spreading}

\begin{figure}[h]
\includegraphics[width=3.0in]{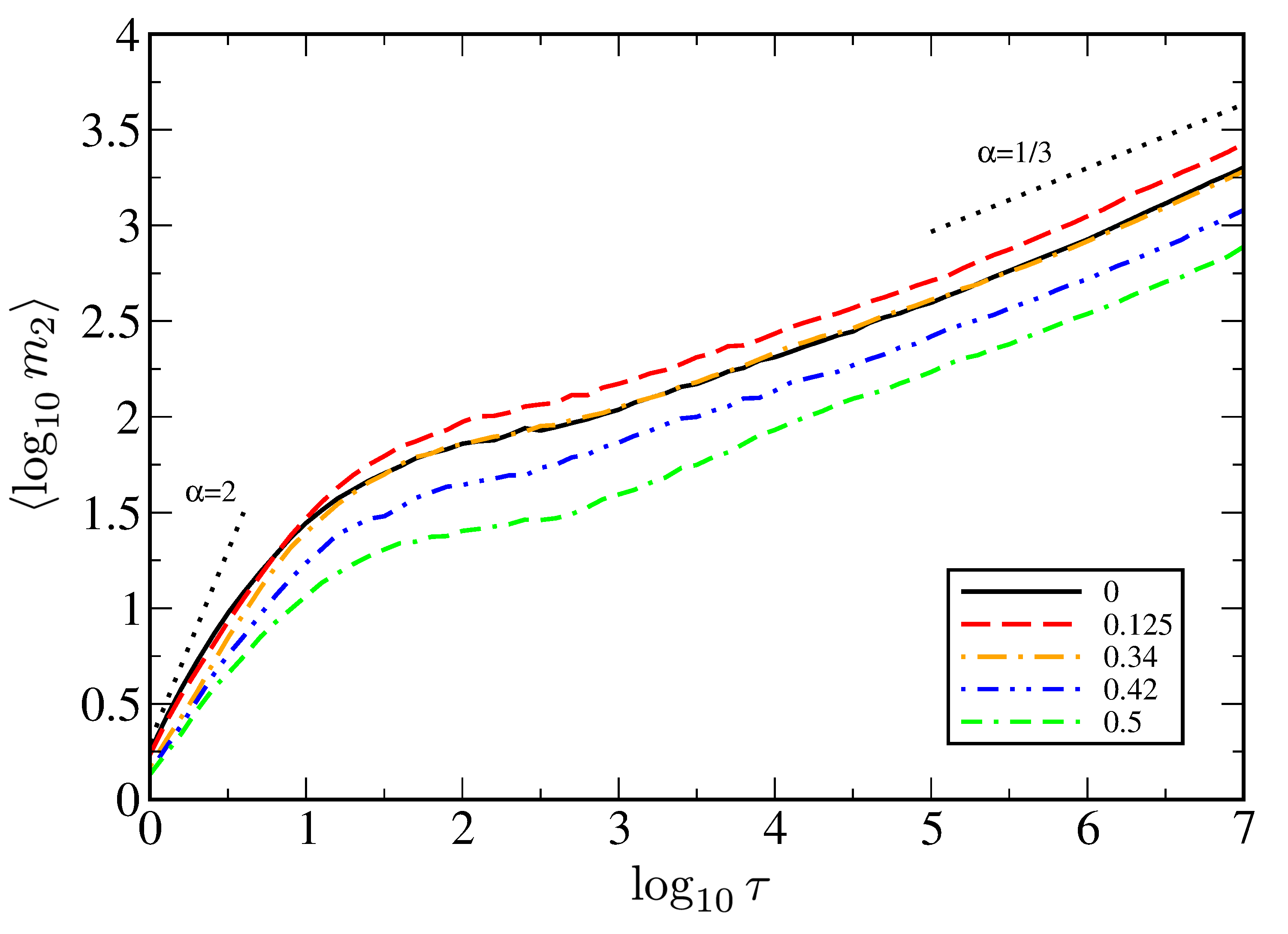}
\vspace{.1cm}
\caption{(Color online) Disorder average over 100 disorder realizations of $\log_{10} m_{2}$ versus $\log_{10} \tau$ 
with disorder strength $W=4$ and $\beta=1$ for
various magnetic fluxes: $q=0,\ 0.125,\ 0.34,\ 0.42,$ and $0.5$.
The two dashed lines guide the eye for $\tau^{\alpha}$ with $\alpha=2$ and $\alpha=1/3$, respectively.}
\label{f:spreading}
\end{figure}

We launch a local excitation in the center of the ladder as an initial wave packet, 
namely, $\psi_{1,l}(\tau=0)=\psi_{2,l}(\tau=0)=\delta_{l,L/2}/\sqrt{2}$, and study the wave packet spreading.
To characterize the wave packet spreading we calculate the second moment $m_{2}\equiv \sum_{l}(l-\overline{l})^2 \Arrowvert \Psi_{l}(\tau)\Arrowvert^2=\sum_{l}(l-\overline{l})^2 \left(|\psi_{1,l}(\tau)|^2+|\psi_{2,l}(\tau)|^2\right) $
with $\overline{l}\equiv\sum_{l}l \Arrowvert \Psi_{l}(0) \Arrowvert^2=\sum_{l}l\left(|\psi_{1,l}(0)|^2+|\psi_{2,l}(0)|^2\right)$.
We use the $\rm SBAB_{2}$ symplectic integrator~\cite{integrators} to evaluate 
the wave function $\Psi_{l}(\tau)$ and therefore $m_{2}$.
The time evolution of $m_{2}$ for various magnetic fluxes is shown in Fig.\ref{f:spreading} for $\beta=1$ and disorder strength $W=4$.

During a first short time regime ($\tau<10$) the wave packet spreads ballistically, 
$m_{2}= g(q)\tau^{\alpha}$ with the exponent $\alpha \simeq 2$.
In this regime, the prefactor $g(q)$ decreases with increasing the magnetic flux.
This is due to the fact that
the largest group velocity of waves in the corresponding linear and clean system decreases with increasing the magnetic flux.
In one-dimensional disordered systems, the localization length is of the same order as the mean free path. 
We can therefore neglect the effect of disorder in this short time regime. Nonlinearity does not affect so much the behavior of the spreading in this regime either. 
The strength of nonlinearity considered here is chosen such as to be in the weak chaos regime~\cite{flachreview}. This implies that the interaction energy of the initial wave packet
is small compared to the band width of the linear wave equation. Therefore the ballistic wave packet spreading up to a distance of the order of the localization length is expected.
Recent studies on the spreading dynamics of interacting bosons in homogeneous lattices show that in one-dimensional lattices, 
for weak interactions, the nonlinearity induced suppression of the expansion velocity is too weak to be observed~\cite{Bloch2}, which is consistent with our observation.
For substantially larger times ($\tau>10^3$) the wave packet exhibits a sub-diffusive behavior.
The second moment of the wave packet grows as $m_{2}= g(q)\tau^{\alpha}$ with the exponent $\alpha \simeq 1/3$, which does not depend on the value of the flux. 
This subdiffusive spreading is caused by chaoticity of the wave packet dynamics which is due to resonances and nonintegrability. 
The exponent $\alpha=1/3$ has been shown to depend solely
on the power of the nonlinear terms, and on the dimensionality of the underlying lattice~\cite{Sergej}.
The presence of synthetic gauge fields does not affect these ingredients, underpinning the universality of the subdiffusive spreading
exponent $\alpha=1/3$.
However, the synthetic gauge field does affect the
prefactor $g(q)$.
For small magnetic fluxes ($0<q < q_c$), the prefactor is enhanced and for big magnetic fluxes ($q_c < q \leq 1/2$) 
the prefactor is reduced. As we will show, this is due to the strong variation of the localization length with changing the magnetic flux.

\begin{figure}[h]
\includegraphics[width=3.4in]{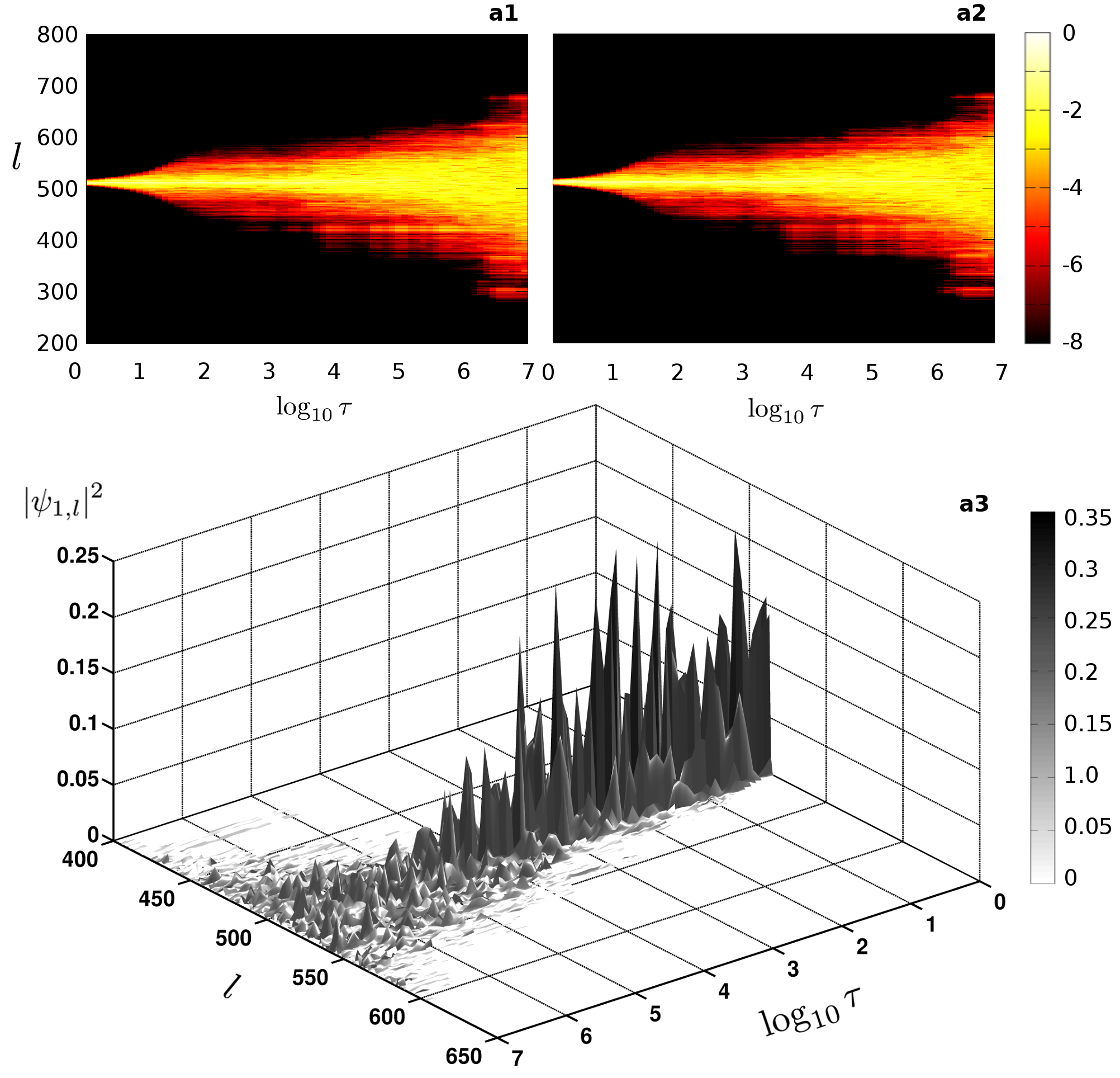}
\vspace{.1cm}
\caption{(Color online) Time evolution of the wave function density for $q=0$. The upper left plot (a1) corresponds to the first leg of the ladder, and
the upper right plot (a2) corresponds to the second leg.  We plot $\log_{10}|\psi_{1(2),l}(\tau)|^{2}$ in a color code versus the ladder coordinate $l$ and time $\tau$.
The three-dimensional plot (a3) at the bottom shows the density evolution $|\psi_{1,l}(\tau)|^{2}$ on a linear scale for the first leg. The second leg plot is very similar and
therefore omitted.
}
\label{f:waveevolutionflux1}
\end{figure}

\begin{figure}[h]
\includegraphics[width=3.1in]{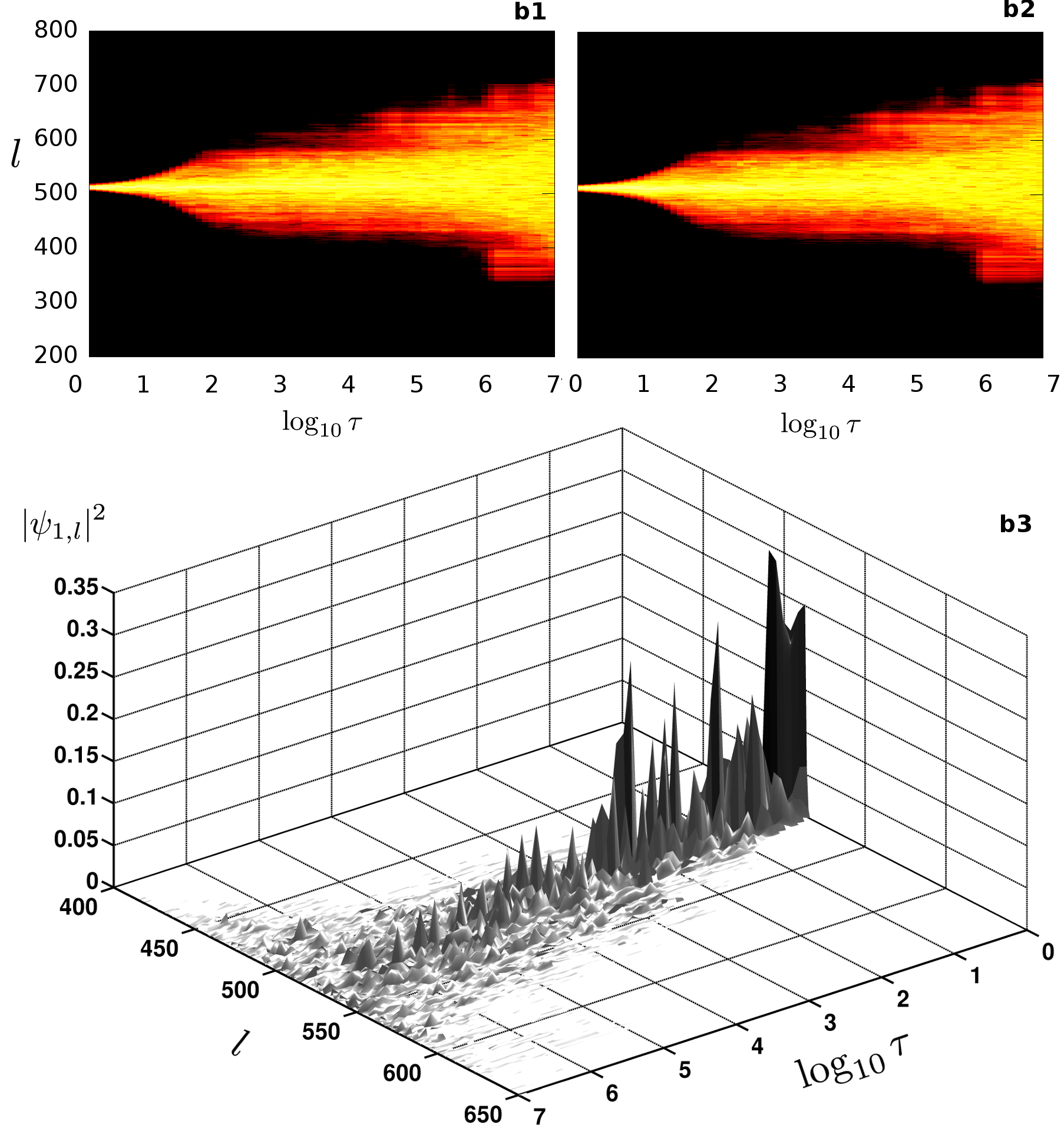}
\vspace{.1cm}
\caption{(Color online) 
Time evolution of the wave function density for $q=0.125$. The upper left plot (b1) corresponds to the first leg of the ladder, and
the upper right plot (b2) corresponds to the second leg.  We plot $\log_{10}|\psi_{1(2),l}(\tau)|^{2}$ in a color code versus the ladder coordinate $l$ and time $\tau$.
The three-dimensional plot (b3) at the bottom shows the density evolution $|\psi_{1,l}(\tau)|^{2}$ on a linear scale for the first leg. The second leg plot is very similar and
therefore omitted. The color map follows Fig.~\ref{f:waveevolutionflux1}.}
\label{f:waveevolutionflux2}
\end{figure}

\begin{figure}[h]
\includegraphics[width=3.1in]{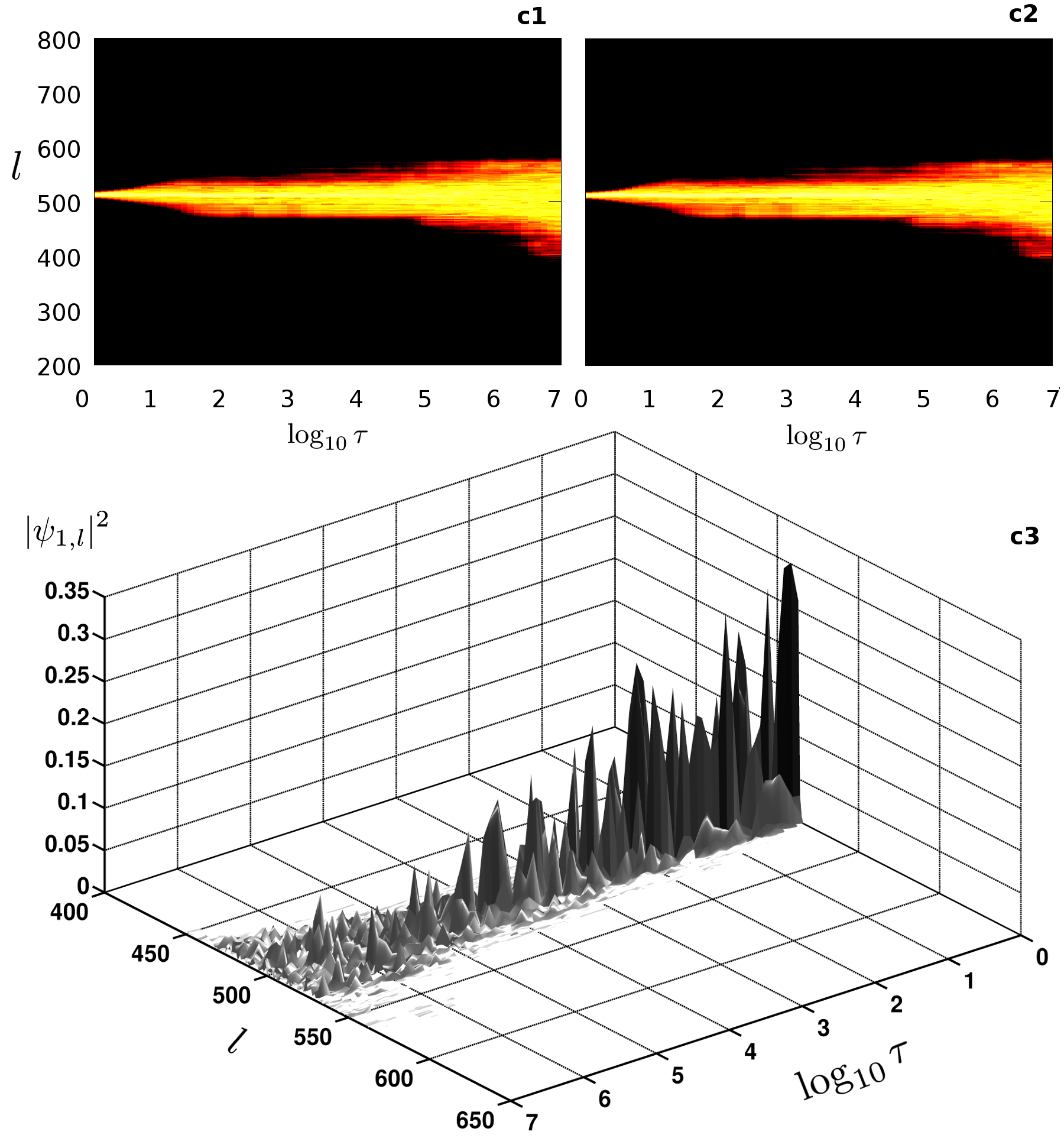}
\vspace{.1cm}
\caption{(Color online) Time evolution of the wave function density for $q=0.5$. The upper left plot (c1) corresponds to the first leg of the ladder, and
the upper right plot (c2) corresponds to the second leg.  We plot $\log_{10}|\psi_{1(2),l}(\tau)|^{2}$ in a color code versus the ladder coordinate $l$ and time $\tau$.
The three-dimensional plot (c3) at the bottom shows the density evolution $|\psi_{1,l}(\tau)|^{2}$ on a linear scale for the first leg. The second leg plot is very similar and
therefore omitted. The color map follows Fig.~\ref{f:waveevolutionflux1}.}
\label{f:waveevolutionflux6}
\end{figure}

In order to get more insight into the details of the dynamics of spreading wave packets, we plot in 
Figs.~\ref{f:waveevolutionflux1}, ~\ref{f:waveevolutionflux2}, ~\ref{f:waveevolutionflux6} 
the space-time dependence of the wave function density along the first ladder leg $|\psi_{1,l}(\tau)|^2$ and the second ladder leg $|\psi_{2,l}(\tau)|^2$
for different magnetic fields.
We observe that the density quickly decays in the core of the wave packet in both legs, leading to a homogeneous spreading without any remnants of Anderson localization
at the original excitation sites, and irrespective of the value of the magnetic field.

\begin{figure}[h]
\includegraphics[width=2.7in]{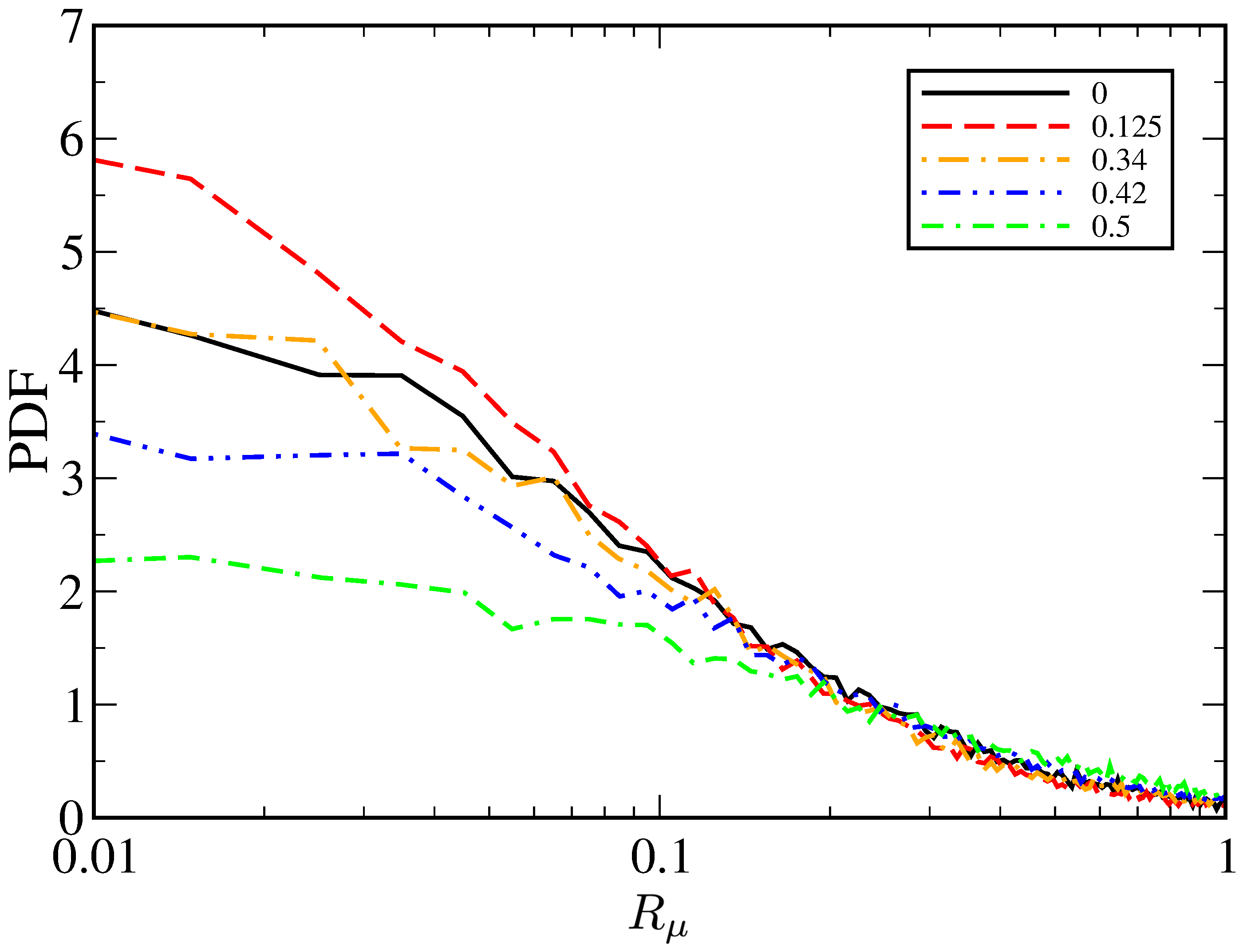}
\vspace{.1cm}
\caption{ The probability distribution function (PDF) of $R_{\mu}$ inside a localization volume at disorder strength $W=4$ for 
various magnetic fluxes: $q=0,\ 0.125,\ 0.34,\ 0.42,$ and $0.5$.
The plot is in log-linear scale.
}
\label{f:statisticsofR}
\end{figure}

Nonlinearity introduces resonances and breaks integrability. 
In the presence of nonlinearity, 
the amplitude of a localized normal mode is modified by a triplet of other excited modes $\vec{\mu}\equiv (\mu_{1},\mu_{2},\mu_{3})$ up to 
the first order in $\beta$ as
\begin{eqnarray}
|c^{(1)}_{\mu}|=\beta\sqrt{n_{\mu_{1}}n_{\mu_{2}}n_{\mu_{3}}} R^{-1}_{\mu,\vec{\mu}}, \quad R_{\mu,\vec{\mu}}\sim \left|\frac{E_{\mu,\vec{\mu}}}{I_{\mu,\mu_{1},\mu_{2},\mu_{3}}} \right|, 
\end{eqnarray}
where $E_{\mu,\vec{\mu}}\equiv E_{\mu}+E_{\mu_{1}}-E_{\mu_{2}}-E_{\mu_{3}}$.
Relevant modes have to reside inside a finite volume of the order of the localization volume of the given mode $\mu$.
To measure the localization volume we use the quantity $\sqrt{12m^{\mu}_2}$~\cite{statisticsofoverlap},
where $m^{\mu}_2\equiv\sum_{l}(l-\overline{l}_{\mu})^{2}||\Phi_{\mu,l}||^2$ and $\overline{l}_{\mu}\equiv\sum_{l}l \Arrowvert \Phi_{\mu,l} \Arrowvert^2$.

Perturbation theory breaks down with the onset of resonances when $\sqrt{n_{\mu}}< |c^{(1)}_{\mu}|$. For simplicity we assume that all modes inside the wave packet
have
the same norm $n$. Then the resonance condition is $\beta n< R_{\mu,\vec{\mu}}$.
For a given normal mode $\mu$, we define $R_{\mu}=\min_{\vec{\mu}} R_{\mu,\vec{\mu}}$, where the minimum is taken inside the corresponding localization volume with $\mu_{1} \neq \mu_{2} \neq \mu_{3} \neq \mu$. 
Collecting $R_{\mu}$ for many $\mu$ and many disorder realizations, we obtain the probability density ${\cal W} (R_{\mu})$ (Fig.~\ref{f:statisticsofR}). 
Following the argument of Ref.\cite{flachreview} the probability, that a given mode has at least one triplet of other modes
with which it is resonant at a given value of $\beta$, is ${\cal P}=\int^{\beta n}_{0} {\cal W}(x) dx$. We denote 
$C={\cal W}(R_{\mu}\rightarrow 0)\neq 0$. 
A bigger (smaller) $C$ indicates a stronger (weaker) resonance. 
We find that for
stronger (weaker) resonance the non-linear spreading evolves faster (slower) (see Fig.~\ref{f:spreading} and Fig.~\ref{f:statisticsofR}).

\begin{figure}[h]
\includegraphics[width=2.8in]{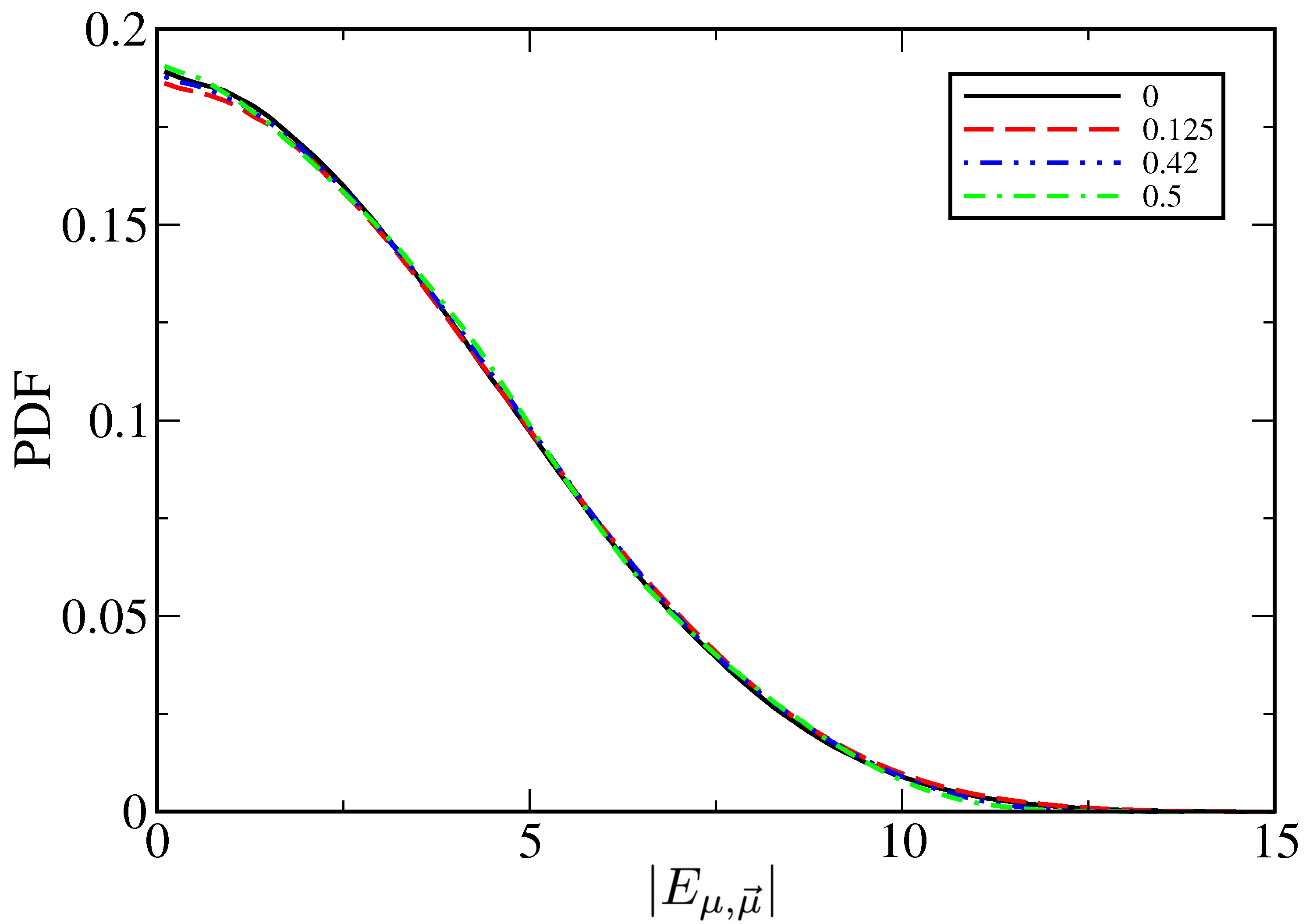}
\vspace{.1cm}
\caption{(Color online) The PDF of $|E_{\mu,\vec{\mu}}|$ inside a localization volume with disorder strength $W=4$ for
various magnetic fluxes: $q=0,\ 0.125,\ 0.42,$ and $0.5$.}
\label{f:quadruplets}
\end{figure}

\begin{figure}[h]
\includegraphics[width=2.8in]{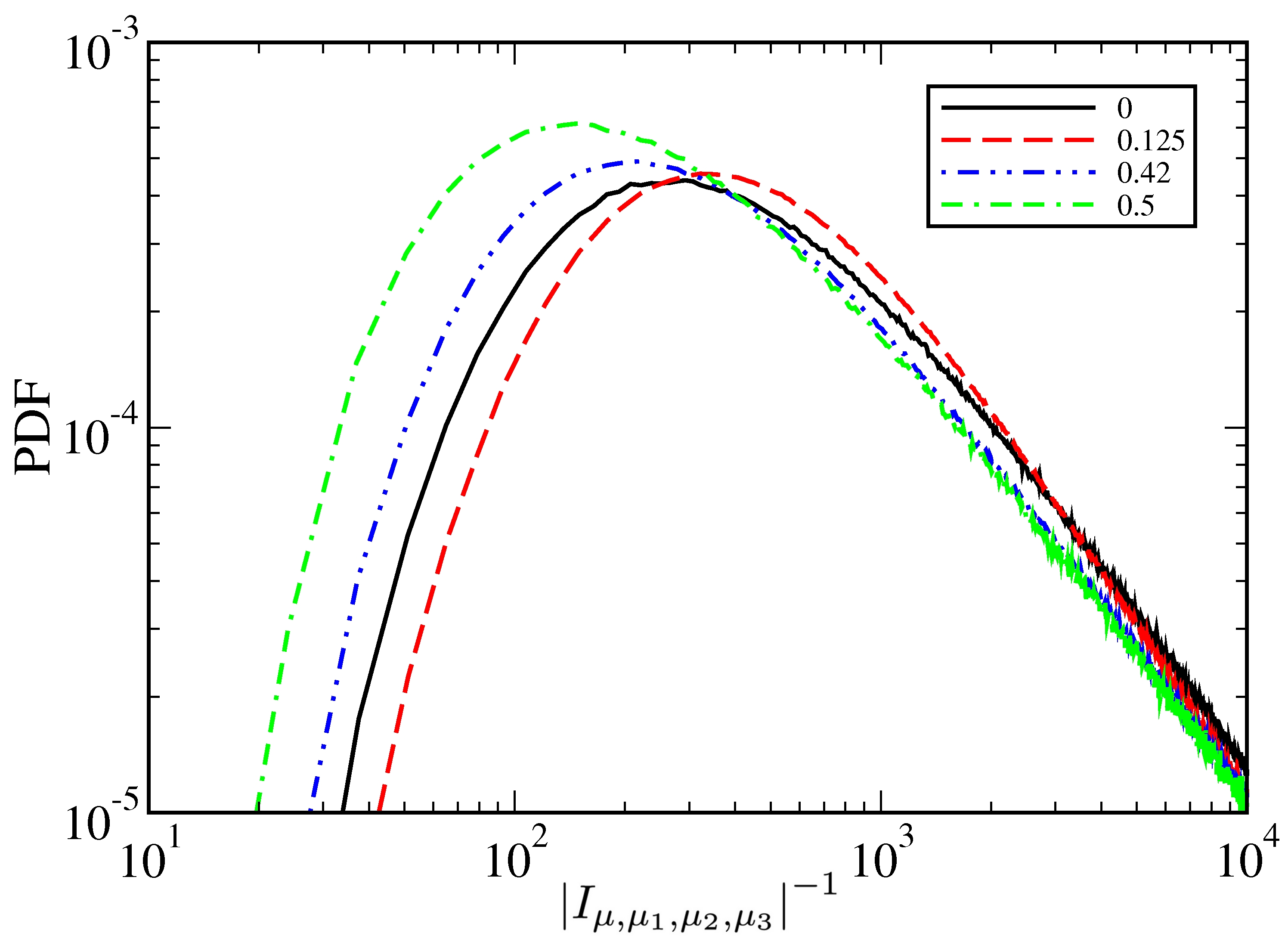}
\vspace{.1cm}
\caption{(Color online) The PDF of $|I_{\mu,\mu_{1},\mu_{2},\mu_{3}}|^{-1}$ inside a localization volume with disorder strength $W=4$ for 
various magnetic fluxes: $q=0,\ 0.125,\ 0.42,$ and $0.5$. The plot is in log-log scale.}
\label{f:inverseoverlap}
\end{figure}

We also study the distribution of $E_{\mu,\vec{\mu}}$ and $|I_{\mu,\mu_{1},\mu_{2},\mu_{3}}|^{-1}$ separately.
We choose a localized eigenstate with energy $E_{\mu}$ and store $E_{\mu,\vec{\mu}}$
for all eigenstates $\mu_{1} \neq \mu_{2} \neq \mu_{3} \neq \mu$ inside the localization volume associated with the state $\mu$.  
We repeat the same procedure for the next eigenstate.  
Collecting these data for many disorder realizations, 
we obtain the statistics of $|E_{\mu,\vec{\mu}}|$. The same method yields the statistics of $|I_{\mu,\mu_{1},\mu_{2},\mu_{3}}|^{-1}$.
For further details we refer to Ref. \cite{statisticsofoverlap}. In
Fig.~\ref{f:quadruplets} and Fig.~\ref{f:inverseoverlap} we plot the distributions of 
$|E_{\mu,\vec{\mu}}|$ and $|I_{\mu,\mu_{1},\mu_{2},\mu_{3}}|^{-1}$
for various magnetic fluxes. The distribution of $|E_{\mu,\vec{\mu}}|$ is insensitive to the magnetic flux and is very close to a Gaussian distribution. 
The reason is that inside a localization volume there are many energy levels, and correlations between them are strong only when their distances are of the order of the mean level spacing.
Since most contributions to $|E_{\mu,\vec{\mu}}|$ come from more distant levels, the distribution of $|E_{\mu,\vec{\mu}}|$ is 
approaching the distribution of sums over independent variables.
The distribution of $|I_{\mu,\mu_{1},\mu_{2},\mu_{3}}|^{-1}$ however depends on the magnetic flux.
For small magnetic fluxes the probability of large overlap integrals becomes smaller, while it is enhanced for larger magnetic fluxes.

\begin{figure}[h]
\includegraphics[width=2.8in]{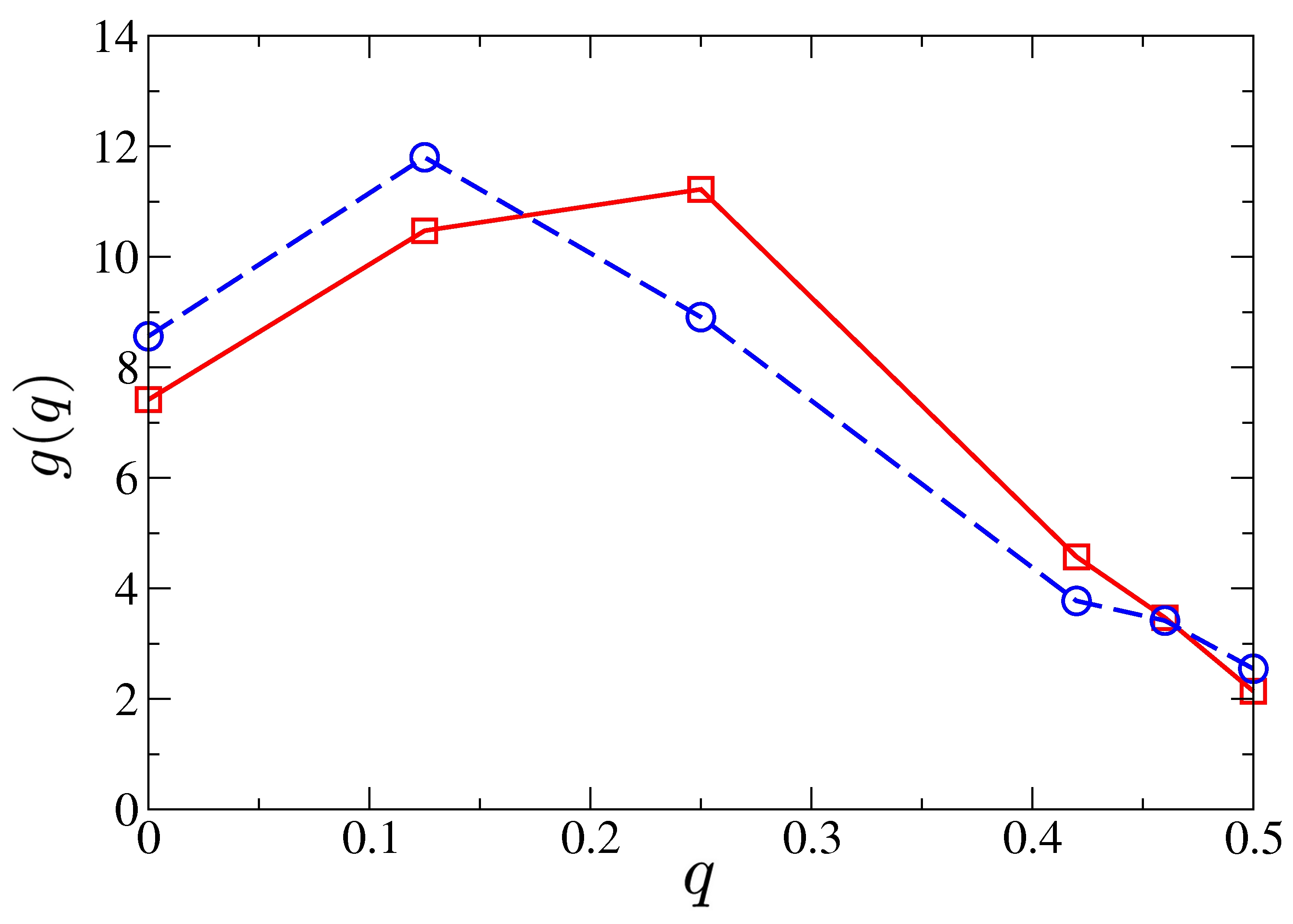}
\vspace{.1cm}
\caption{(Color online) The blue dashed line shows the theoretical prediction of the prefactor $g(q)$ given by Eq.(\ref{coefficent}) with choosing $|\log_{10} C'|=0.91$.
The red solid line shows the coefficient extracted from the numerical data in the subdiffusive regime for various magnetic fluxes: $q=0,\ 0.125,\ 0.25,\ 0.42,\ 0.46,$ and $0.5$.
}    
\label{f:coefficent}
\end{figure}

In order to understand the variation of the prefactor $g(q)$ with the magnetic flux,
we adopt the arguments of Refs.~\cite{Sergej,flachreview,intermediatestatistics,statisticsofoverlap}. 
The main idea is to assume that a part of the normal modes in the wave packet is chaotic due to the nonlinearity.
If the overlap $I_{\mu,\mu_{1},\mu_{2},\mu_{3}}$ does not fluctuate strongly, we can replace it by its average $\langle I\rangle$  
and replace $c_{\mu_{1}}c^{\ast}_{\mu_{2}}c_{\mu_{3}}$ by $n^{3/2}$ in Eq.~(\ref{NLEOM}), where $\langle I\rangle$ is the average overlap in the wave packet. 
Then we consider a Langevin type equation of motion of a mode $\hat{\mu}$ outside of the wave packet which however resides in a close neighborhood to the wave packet: 
\begin{eqnarray}
\label{SEOM}
i\dot{c}_{\hat{\mu}}\approx E_{\hat{\mu}}c_{\hat{\mu}}+\beta \langle I\rangle \overline{\xi}^{3} n^{3/2} {\cal P}(\beta n) f(\tau),  
\end{eqnarray}
where $f(\tau)$ is generated by the chaotic dynamics of wave packet modes and is assumed to be an uncorrelated white noise, 
$\langle f(\tau)f(\tau')\rangle=\delta(\tau-\tau')$. Here $\overline{\xi}\equiv\int \rho(E)\xi(E) dE$ is the average of the localization length over all the modes.
Such chaotic dynamics was confirmed in recent quantitative studies~\cite{chaos}.
From the numerical data we observe that $C/\overline{\xi}$ is roughly independent of the flux, namely $C \sim \overline{\xi}.$ 
In the weak chaos regime $C\beta n<1$, ${\cal P}\sim C\beta n$, one obtains 
\begin{eqnarray}
\label{coefficent}
m_{2}\approx C'\beta^{4/3}\langle I\rangle^{2/3}\overline{\xi}^{8/3}\tau^{1/3},
\end{eqnarray}
where $C'$ is a constant which does not depend on any physical parameter~\cite{footnotenormalmodes}. 
Fig.~\ref{f:coefficent} shows the comparison of the coefficient in
Eq.(\ref{coefficent}) and
the coefficient $g(q)$ extracted from the numerical data of Fig.~\ref{f:spreading} for various magnetic fluxes. 
The unknown coefficient $C'$ is fitted to be $|\log_{10} C'|=0.91$. 
The prediction of Eq.(\ref{coefficent}) matches the numerical data reasonably well. 
In particular the observed increase of the prefactor for small magnetic fluxes and the subsequent decrease for larger magnetic fluxes are very well reproduced.

\section{Summary}

We studied the subdiffusive spreading of nonlinear waves in a one-dimensional disordered lattice in the absence of time-reversal symmetry.
Our results show that the much-debated weak chaos subdiffusion law with exponent $\alpha=1/3$ keeps its universality even in the presence
of synthetic gauge fields. Their main impact is to change the localization length. For small magnetic fluxes, the losing of time-reversal symmetry leads to a suppression of backscattering
and an increase of the localization length. Consequently the prefactor $g$ of the subdiffusive spreading law increases. For large magnetic fluxes, the spectrum of the linear wave equations
opens a gap filled with Lifshitz-tail-like localized states. In this regime, the localization length is reduced. It follows that the prefactor $g$ is decreasing.
A theoretical estimate of the dependence of $g$ on the magnetic flux yields good agreement with numerical data. 

It would be also interesting to extend this study to the regime of strong chaos ($\alpha=1/2$)~\cite{strongchaos1,strongchaos2}
and two-dimensional disordered lattices.
 
\begin{acknowledgments}
We thank D. M. Basko, M. M\"{u}ller, J. D. Bodyfelt, and A. M. Mateo for useful discussions.
\end{acknowledgments}

\appendix


\begin{thebibliography}{99}
\bibitem{footnoteelectron} In most electronic lattice systems it would take a large magnetic field $B \sim 10^5$T to 
generate a flux per lattice cell which is of the order of a flux quantum.
\bibitem{footnotegaugefixing}
One can also choose other gauges which may break the lattice translation symmetry of the Hamiltonian along horizontal direction. 
However the magnetic flux is transnational invariant, therefore one can always define a magnetic translator  
which commutes with the Hamiltonian~\cite{Zak}.
\bibitem{footnotenormalmodes}
Notice that here we calculate $m_{2}$ in the normal mode space and in the numerical simulation 
$m_{2}$ is evaluated in the real space. The two are closely related and show the similar behavior.
\end{thebibliography}
\end{document}